# Evidence of Systematic Bias in 2008 Presidential Polling

(preliminary report)

**Leonard Adleman**
**Department of Computer Science**
**University of Southern California**

**Mark Schilling**
**Department of Mathematics**
**California State University, Northridge**

**Abstract**

Political polls achieve their results by sampling a small number of potential voters rather than the population as a whole. This leads to “sampling error” which most polling agencies dutifully report. But factors such as nonrepresentative samples, question wording and nonresponse can produce *non*-sampling errors. While pollsters are aware of such errors, they are difficult to quantify and seldom reported. When a polling agency, whether by intention or not, produces results with non-sampling errors that systematically favor one candidate over another, then that agency’s poll is biased. We analyzed polling data for the (on-going) 2008 Presidential race, and though our methods do not allow us to identify which agencies’ polls are biased, they do provide significant evidence that some agencies’ polls are.

We compared polls produced by major television networks with those produced by Gallup and Rasmussen. We found that, taken as a whole, polls produced by the networks were significantly to the left of those produced by Gallup and Rasmussen.

We used the available data to provide a tentative ordering of the major television networks’ polls from right to left. Our order (right to left) was: FOX, CNN, NBC (which partners with the Wall Street Journal), ABC (which partners with the Washington Post), CBS (which partners with the New York Times). These results appear to comport well with the commonly held informal perceptions of the political leanings of these agencies.

We also compared tracking polls produced by Gallup, Rasmussen, Hotline/FD, and the Daily KOS. Here again we found significant evidence of bias. Most notably, the Rasmussen and the Gallup polls were significantly to the right of the Daily KOS poll. A detailed analysis of the Gallup and Rasmussen polls also suggested the likelihood of short-term bias.

Our findings are preliminary, but given the importance of polling in America, they make a case for further research into the causes of and remedies for polling bias.

### *Introduction*

The influence of opinion polls on political matters has steadily grown in recent years; indeed, poll results now commonly affect political discourse, impact policy decisions and determine campaign strategies. The current election season has spawned a plethora of presidential polls and a number of websites have emerged that combine the information from many polls into a single "state of the race" report that is updated daily.

Unfortunately, polling is not a perfect means of determining the "state of the race". Polls achieve their results by sampling a small number of potential voters rather than the whole population. This leads to "sampling error". For most polls, this sampling error ranges from 2-5% and most polling agencies are diligent in reporting this error.

But polls are also subject to *non*-sampling errors. Pollsters are aware that many factors such as nonrepresentative samples, question wording and question ordering, nonresponse and interviewer bias can affect poll results. The best polling agencies try to mitigate the impact of these factors. When polls are reported in the media, however, non-sampling errors are almost never mentioned. If pollsters were entirely successful in eliminating non-sampling errors, then they could sensibly be ignored by the media and poll consumers. A main purpose of this paper is to provide *prima facie* evidence that pollsters are *not* successful. Our analysis shows that in this election cycle some polls have exhibited a left/right political bias.

We collected data from the following sources:

**Table 1: Data Sources**

| Data | Source |
|---|---|
| Real Clear Politics Average | http://www.realclearpolitics.com |
| Rasmussen Tracking Poll | http://www.rasmussenreports.com |
| Gallup Tracking Poll | http://elections.nytimes.com<br>http://pollingreport.com |
| All other data | http://www.realclearpolitics.com |

Statistically, bias refers to the tendency for an estimator to produce results that are not centered at the target parameter; that is, the estimates tend to be consistently too high or too low. In presidential polling, however, the target is unknown (except on the day of the election), and it is constantly moving as people's opinions change over time. In addition, a precise description of the sampling scheme is typically not made public by the polling agency. For these reasons it is not possible for us to determine whether bias is present for an individual agency's poll. However, it is possible with the data available to investigate whether there is a *bias* of one agency's poll relative to another agency's poll.

To analyze the bias of one agency's poll relative to another, we look at each day on which both agencies' polls report. For each report, we calculate its *spread*, that is, the difference between the percent of respondents supporting Barack Obama and the percent of respondents supporting John McCain. For example, if on a given day Poll A reports Obama with 48% and McCain with 43%, then the Poll A spread is 5%. If, on the same day, Poll B reports Obama with 46% and McCain with 47%, then the Poll B spread is -1%. We then calculate the *difference* of Poll A relative to Poll B for that day, that is, Poll A's spread minus Poll B's spread. In this example, the difference of Poll A relative to Poll B is 5%-(-1%) = 6%. If the difference is positive we refer to it as "Pro-Obama", if negative as "Pro-McCain". We estimate the bias of Poll A relative to Poll B as the average of the differences of Poll A relative to Poll B over as many days as we have data. We use the collection of all differences of Poll A relative to Poll B to establish whether any observed systematic bias of Poll A relative to Poll B is statistically significant. It is important to note that if there is a Pro-Obama bias of Poll A relative to Poll B, then there is a Pro-McCain bias of Poll B relative to Poll A; if there is a Pro-McCain bias of Poll A relative to Poll B, then there is a Pro-Obama bias of Poll B relative to Poll A.

### *Polls by the Major Television Networks*

We analyzed the polling data for ABC, CBS, CNN, FOX, and NBC during the period from March 12, 2008 through September 29, 2008. Not only is each of these networks of great importance in its own right; their polls are a suitable surrogate for main-stream media polls since the networks often collaborate with major newspapers when polling. In particular, ABC collaborates with the Washington Post, CBS with the New York Times, and NBC with the Wall Street Journal.

There were 43 individual reports by major networks during the period of our investigation. We compared these reports to those of the daily tracking poll conducted by Gallup. Of the 43 network reports, 42 were reported on days that Gallup also reported. Our findings are in Table 2 below:

**Table 2: Major Network Polls relative to the Gallup Tacking Poll**

| Bias of Major Networks relative to | Number of reports | Number of times Major Networks | Number of times Major Networks |
|---|---|---|---|

| Gallup | considered | Pro-Obama relative to Gallup | Pro-McCain relative to Gallup |
|---|---|---|---|
| 2.67% Pro-Obama | 42 | 31 | 6 |

The 2.67% Pro-Obama bias of the major networks relative to Gallup is suggestive, but it is important to keep in mind that it is equivalent to say that there is a 2.67% Pro-McCain bias of Gallup relative to the major networks. It does not necessarily mean that the major networks are biased relative to the underlying truth, nor does it necessarily mean that Gallup is either. However, it does suggest that one or the other and perhaps both are biased relative to the underlying truth. Informally, the major networks and Gallup might be compared to two scales, one of which on average reads 2.7 pounds higher than the other, but the user does not know which, if either, is correct.

Note also that the major networks were Pro-Obama relative to Gallup 31 times and Pro-McCain relative to Gallup only 6 times. This evidence for relative bias is highly significant statistically, as the probability of a fair coin achieving 31 or more of one particular outcome in 37 flips is only .00004.

We also compared the major networks polls with the daily tracking poll conducted by Rasmussen. Of the 43 network reports, 29 were reported on days that Rasmussen also reported. The results are similar to those seen relative to Gallup, but the Pro-Obama bias relative to Rasmussen is less pronounced:

**Table 3: Major Network Polls relative to the Rasmussen Tacking Poll**

| Bias of Major Networks relative to Rasmussen | Number of reports considered | Number of times Major Networks Pro-Obama relative to Rasmussen | Number of times Major Networks Pro-McCain relative to Rasmussen |
|---|---|---|---|
| 1.48% Pro-Obama | 29 | 19 | 8 |

We conducted a formal statistical analysis of the mean relative biases between the major network polls and the Gallup and Rasmussen polls. The Gallup tracking poll used a 5-day moving average until June 8, 2008 and thereafter used a 3-day moving average. The Rasmussen tracking poll used a 3-day moving average from its inception on June 8, 2008. In order to avoid using results that were not independent, we avoided using poll reports with reporting dates less than 5-days apart during the period up to June 8, 2008, and less than 3-days apart thereafter. With this constraint, we determined a maximal set of 24 individual network reports that comprised our “Major Media poll”.

We compared the Major Media poll with the daily tracking poll conducted by Gallup by means of a two-sided paired t-test applied to the sets of daily differences between the Major Media and Gallup polls, testing the hypothesis of no difference in mean spread. Of the 24 individual reports, 23 were on days that Gallup also reported. Our findings are given below:

**Table 4: Major Media Poll relative to the Gallup Tacking Poll**

| Bias of Major Media relative to Gallup | p-value | Number of Reports considered | Number of times Major Media Pro-Obama relative to Gallup | Number of times Major Media Pro-McCain relative to Gallup |
|---|---|---|---|---|
| 2.87% Pro-Obama | .0026 | 23 | 17 | 2 |

We also compared the Major Media poll with the daily tracking poll conducted by Rasmussen. Of the 24 individual reports, 23 were on days that Rasmussen also reported.

**Table 5: Major Media Poll relative to the Rasmussen Tacking Poll**

| Bias of Major Media relative to Rasmussen | p-value | Number of Reports Considered | Number of times Major Media Pro-Obama relative to Rasmussen | Number of times Major Media Pro-McCain relative to Rasmussen |
|---|---|---|---|---|
| 1.94% Pro-Obama | .02 | 16 | 11 | 3 |

Tables 4 and 5 above show that during our study period the Gallup and Rasmussen polls produced results that were typically more Pro-McCain than the Major Media poll, and that these results were very unlikely to be chance occurrences. The magnitude of the bias (2.87%) of the Major Media relative to Gallup is comparable to the magnitude of often-cited sampling errors, and indicates that bias is a non-trivial component of polls that report the state of the presidential race. Given the prevailing belief that polls can affect election outcomes, it is important that the possibility of such bias be generally acknowledged when polls are reported.

We next looked at each TV network individually and compared its poll with those of Gallup and Rasmussen. Because each network produced a small number of reports during our period of investigation one might expect the statistical significance of any bias to be small. In fact, however, several comparisons *were* statistically significant at standard levels. The following table records our findings:

**Table 6: Individual TV Network Polls relative to the Gallup Tacking Poll and the Rasmussen Tracking Poll**

| | Bias relative to Gallup Tracking | Bias relative to Rasmussen Tracking |
|---|---|---|
| FOX | 1.33%<br>Pro-Obama<br>(6, .14) | 0.00%<br><br>(5, 1) |
| CNN | 1.78%<br>Pro-Obama<br>(9, .18) | 0.67%<br>Pro-Obama<br>(6, .61) |
| ABC/Washington Post | 2.43%<br>Pro-Obama<br>(7, .02*) | 2.40%<br>Pro-Obama<br>(5, .24) |
| NBC/Wall Street Journal | 2.57%<br>Pro-Obama<br>(7, .02*) | 1.80%<br>Pro-Obama<br>(5, .10) |
| CBS/New York Times | 4.08%<br>Pro-Obama<br>(13, .02*) | 2.25%<br>Pro-Obama<br>(8, .13) |

(Values in parentheses are ($N$, $p$) = # reports, p-value for the hypothesis that the bias of TV network relative to tracking poll is zero. * indicates $p < .05$)

During the period of our analysis no major TV network had a Pro-McCain bias relative to either Gallup or Rasmussen. Or put alternatively, neither Gallup nor Rasmussen had a Pro-Obama bias relative to any major TV network.

We then compared each network poll to the Real Clear Politics Average. We could not discover sufficient information about how Real Clear Politics calculates its average to justify a formal statistical analysis, however, our results are shown in the following table:

**Table 7: Individual TV Network Polls relative to the Real Clear Politics Average**

| | Number of times Pro-Obama relative to RCP | Number of times Pro-McCain relative to RCP | Bias Relative to RCP |
|---|---|---|---|
| FOX | 3 | 4 | 0.40% Pro-McCain |
| CNN | 6 | 3 | 0.67% Pro-Obama |
| NBC/Wall Street Journal | 5 | 2 | 1.08% Pro-Obama |
| ABC/Washington Post | 6 | 1 | 2.44% Pro-Obama |
| CBS/New York Times | 11 | 2 | 2.83% Pro-Obama |

From the data obtained, we ranked the TV networks in order of their bias relative to Gallup, Rasmussen and Real Clear Politics.

**Table 8: Rankings of Individual TV Network Polls**

| | Rank relative to Gallup Tracking | Rank relative to Rasmussen Tracking | Rank relative to Real Clear Politics |
|---|---|---|---|
| FOX | 1 | 1 | 1 |
| CNN | 2 | 2 | 2 |
| ABC/Washington Post | 3 | 5 | 4 |
| NBC/Wall Street Journal | 4 | 3 | 3 |
| CBS/New York Times | 5 | 4 | 5 |

Notice that relative to Gallup, Rasmussen and RCP the order of the rankings of FOX, CNN, NBC/Wall Street Journal, and CBS/New York Times is the same. However, ABC/Washington Post has different positions relative to the NBC/Wall Street Journal, and CBS/New York Times. We were led to the following ordering (from right to left): FOX, CNN, NBC/Wall Street Journal, ABC/Washington Post, CBS/New York Times.

Relative to the Real Clear Politics Average, the CBS/New York Times poll had a 2.83% Pro-Obama bias, while the FOX poll had a 0.40% Pro-McCain bias. Although the polling days for the CBS/New York Times poll and the FOX poll were different, it seems reasonable to place FOX 3.23% (=2.83%-(-0.40%)) to the right of the CBS/New York Times. Using this metric we produced the following graph which estimates the relative distances along the left/right spectrum of the network polls:

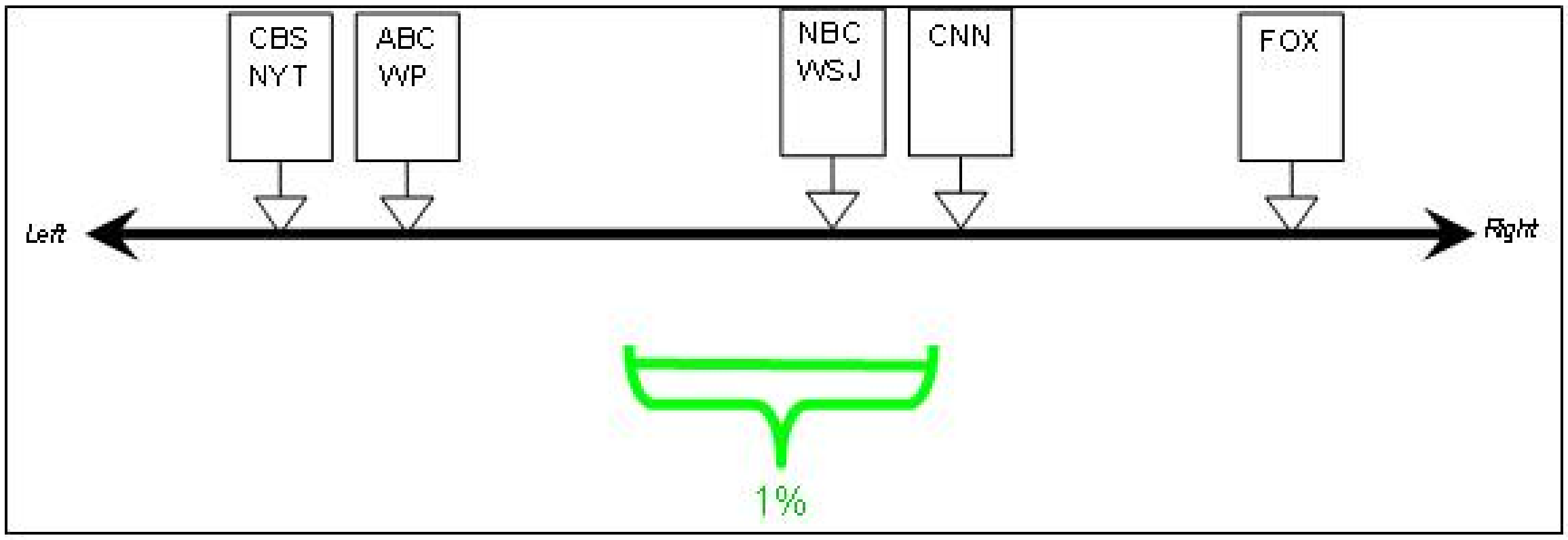

**Figure 1: Estimated Left/Right Ordering of Network Polls with Estimated Distances for the Period 3/12/08 to 9/29/08**

### *Tracking Polls*

Our analysis of tracking polls involves four polls that have provided daily results during the 2008 presidential campaign. The Gallup tracking poll has been providing data since March 12, 2008, the Rasmussen poll since June 8, 2008, Hotline/FD since September 8, 2008 and the Daily KOS since September 11, 2008. With the exception of the Gallup tracking poll which used a 5-day moving average prior to July 9, 2008, all results of these polls are based on a 3-day moving average. Figure 2 shows the spreads for each of these polls during the period of September 9, 2008 to September 29, 2008:

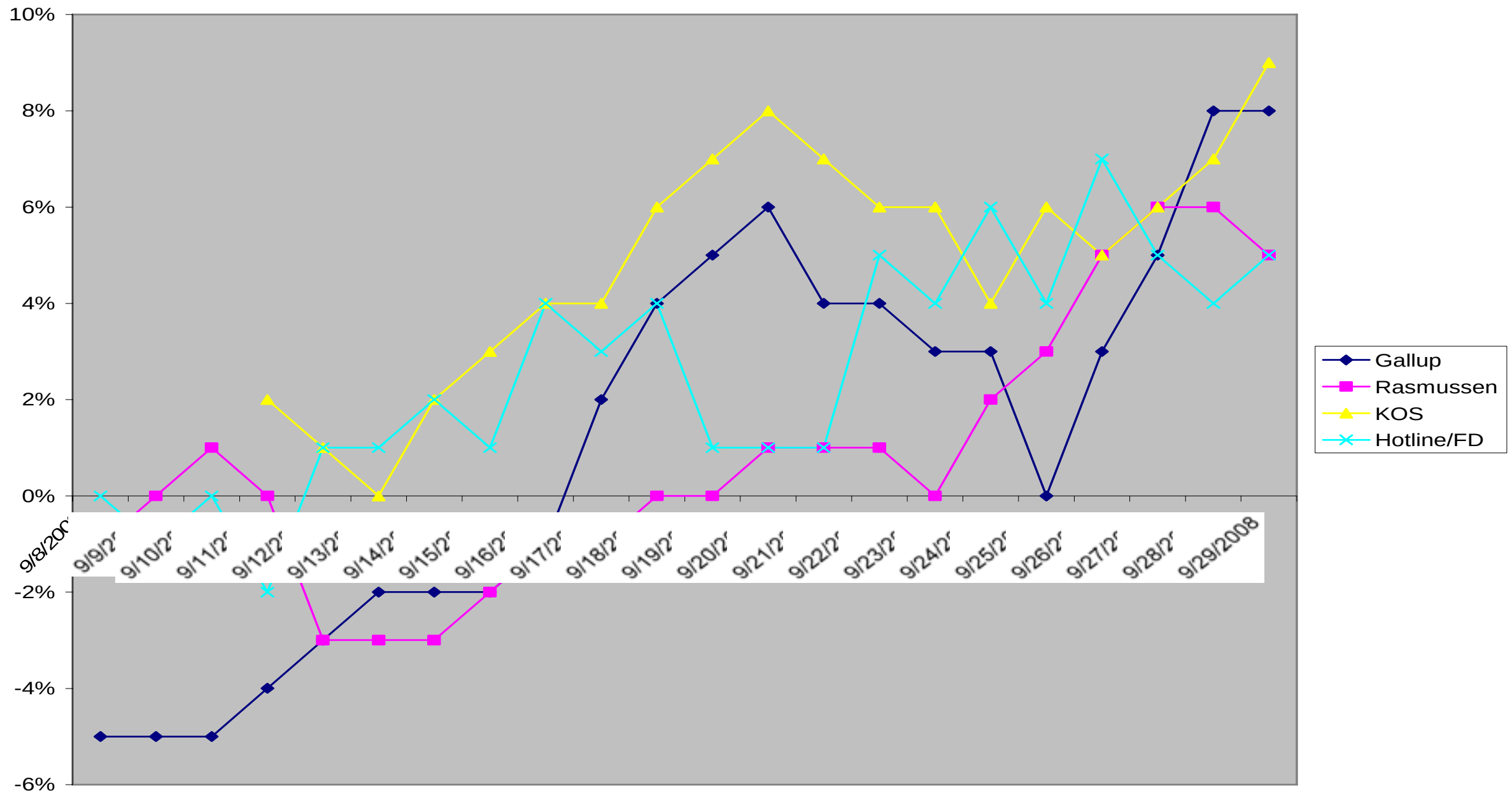

**Figure 2: Spreads for Four Tracking Polls**

One can see that the Daily KOS poll produced no negative spreads (i.e., McCain was never ahead), and produced larger spreads in favor of Obama than the other three polls on most days that the Daily KOS poll was in operation. This suggests the possibility of a Pro-Obama bias in the Daily KOS poll relative to the other polls. To address whether there is significant evidence of bias, we performed two analyses of variance on this data.

First, since the data in each tracking poll contains serial dependence due to the use of a three day moving average, we extracted and used only the results from every third day. Next we detrended the data by subtracting the RCP Average margin. For completeness, the next table shows the bias of Real Clear Politics relative to each of the tracking polls.

**Table 9: Real Clear Politics Relative to the Tracking Polls**

| | |
|---|---|
| Bias of RCP Relative To Daily KOS | 3.89% Pro-McCain |
| Bias of RCP Relative To Hotline/FD | 1.76% Pro-McCain |
| Bias of RCP Relative to Gallup | 0.64% Pro-McCain |
| Bias of RCP Relative to Rasmussen | 0.01% Pro-McCain |

After detrending, we assumed that each tracking poll's results represent independent observations from a common distribution. We used a one-way analysis of variance to test the hypothesis that there is no difference between the mean values of the tracking polls. The results show highly significant evidence ($p$ = .002) that there *is* a difference, i.e., that there was bias present during this time period.

Since the data for all four polls were taken on the same days, we also applied a two-way analysis of variance, where the days represent the blocks, and found similar and significant results ($p$ = .0076).

Finding such significant evidence with a rather small amount of data is an indication that the magnitude of the bias is large in comparison to the day to day variation in each poll's data. We conclude that there is good support for the notion that bias was present among the tracking polls over the period under study.

A multiple comparisons analysis shows a statistically significant difference between Daily KOS and Gallup, and between Daily KOS and Rasmussen at $\alpha$ = .05, indicating the strong possibility of bias.

The Figure above suggests that the bias (if any) of Gallup relative to Rasmussen was small during the period of September 9, 2008 to September 29, 2008. However, if we consider the entire period for which both polls used a 3-day moving average there is evidence that during some periods there was relative bias between Gallup and Rasmussen.

Figure 3 shows the differences of the Gallup tracking poll relative to the Rasmussen tracking poll over the entire period for which both polls used a 3-day moving average:

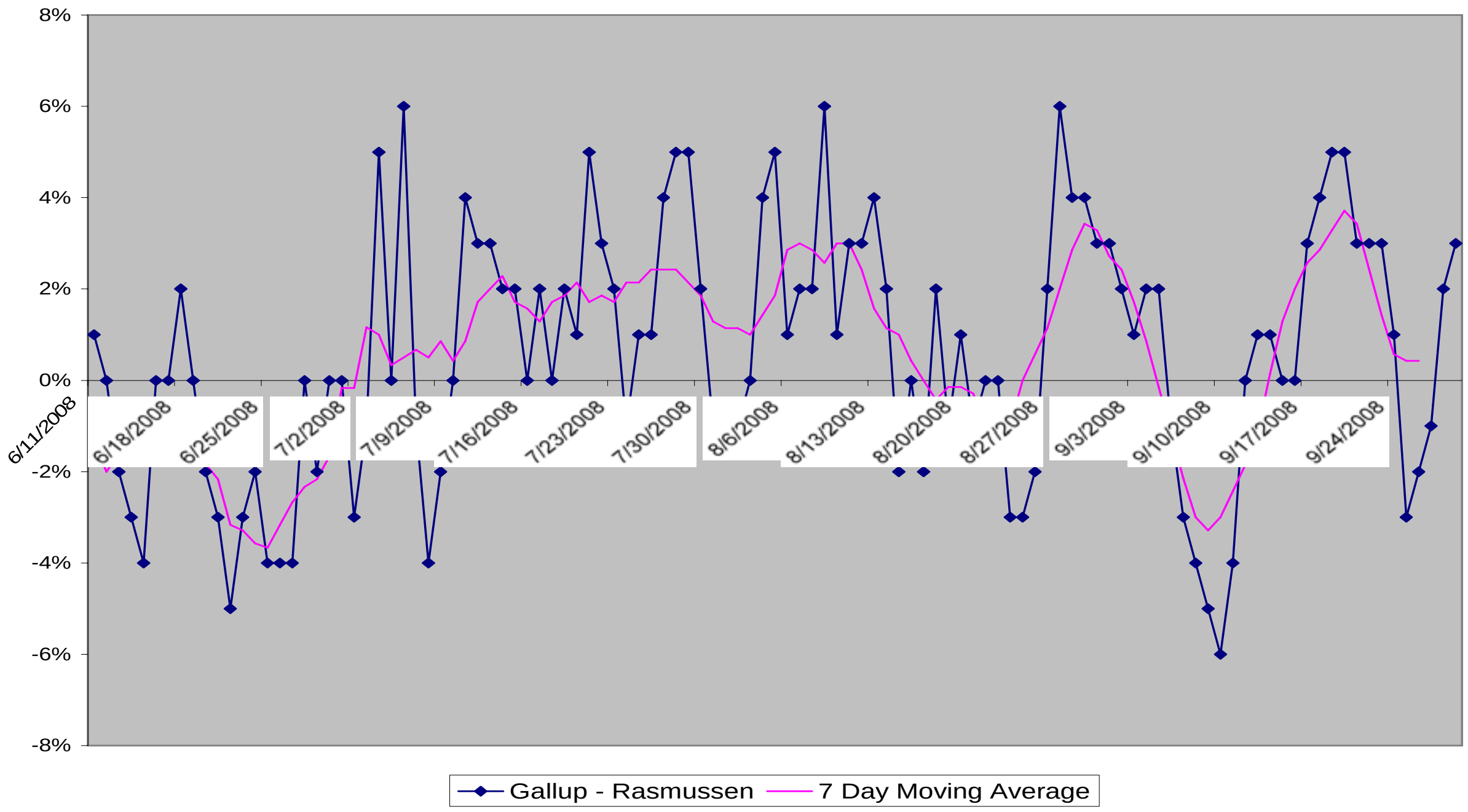

**Figure 3: Differences of Gallup Tracking Poll Relative to Rasmussen Tracking Poll**

Note that the differences tended rather consistently negative for nearly a month, after which the differences tended to the positive side for an even longer period. That is, Gallup tended Pro-McCain relative to Rasmussen for nearly a month, and then tended Pro-Obama relative to Rasmussen for at least a month. To assess the likelihood that this pattern was accidental, we again extracted and analyzed the results from every third day to assure independence:

**Table 10:**
**Differences of Gallup relative to Rasmussen**

| DATE | Difference of Gallup relative to Rasmussen |
|---|---|
| 6/11008 | 1% |
| 6/14/2008 | –3% |
| 6/17/2008 | 0% |
| 6/20/2008 | –2% |
| 6/23/2008 | –3% |
| 6/26/2008 | –4% |
| 6/29/2008 | –2% |
| 7/2/2008 | –3% |
| 7/8/2008 | –4% |
| 7/11/2008 | 4% |
| 7/14/2008 | 2% |
| 7/17/2008 | 2% |
| 7/20/2008 | 1% |
| 7/23/2008 | 2% |
| 7/26/2008 | 1% |
| 7/29/2008 | 5% |
| 8/1/2008 | -1% |
| 8/4/2008 | 4% |
| 8/7/2008 | 2% |
| 8/10/2008 | 1% |
| 8/13/2008 | 4% |
| 8/16/2008 | 0% |
| 8/19/2008 | -1% |
| 8/22/2008 | 0% |
| 8/25/2008 | -3% |
| 8/28/2008 | 6% |
| 8/31/2008 | 3% |
| 9/3/2008 | 1% |
| 9/6/2008 | -1% |
| 9/9/2008 | -5% |
| 9/12/2008 | 0% |
| 9/15/2008 | 0% |
| 9/18/2008 | 4% |
| 9/21/2008 | 3% |
| 9/24/2008 | 1% |
| 9/27/2008 | -1% |

The shadings indicate blocks of consecutive values in which the relative difference is negative (light shading) and blocks of consecutive values in which the relative difference is positive (dark shading). If in fact there is no tendency towards bias that persists over independent polls, we would expect to see a random pattern of positive, negative and some zero differences. The runs we see above for the period between mid-June and mid-September are suggestive of a nonrandom pattern.  In particular, there is a period of 12 days during which 11 values are positive.

We performed a permutation test by means of simulation to assess the statistical significance of the period in which the difference was positive for 11 out of 12 days. In each simulation we created a random permutation of the percentages given above. In only 31 out of the 100,000 simulations we ran did we find a period in which the differences were positive at least eleven out of twelve consecutive days. Although we acknowledge that this is a *post hoc* analysis, the smallness of the significance probability that we found ($p = .00031$) still provides evidence that, for at least some portion of the study period, there was relative bias between the Gallup and Rasmussen polls. It is our impression that both of these polling agencies are diligent in their efforts to combat bias. Our findings illustrate that even under these circumstances, bias can arise. Bias is not an anomaly in polling, it is the natural state.

### ***Conclusion***

We have presented a *prima facie* case for the existence of bias in political polling. Given the importance of polling in the modern day election process, it is important that this issue be addressed. We have some suggestions.

1. Pollsters, in addition to declaring the degree of sampling error in their polls, should make declarations regarding the presence of non-sampling errors and in particular the fact that these may include bias. We recognize that such errors are difficult to identify and quantify. Nonetheless, a pollster should reveal what it construes as possible sources of error and what steps it has taken to mitigate their impact.

2. News disseminating organizations that report polling results should include information about sampling errors, non-sampling errors and bias. They should indicate what pollsters claim about their results, but should also make independent judgments as well.

3. Pollsters should reveal more about the internal workings of their polls. The need to protect proprietary methods should be balanced with the value of enabling independent scrutiny of the strengths and weaknesses of the methods used.

4. More research should be conducted into the nature of bias, its prevalence, its impact, its causes and its remedies. We might gain insight by considering approaches taken to avoid researcher bias in science: control groups, double blind studies, statistical requirements for significance, eschewing of anecdotal evidence.

5. Bias is a common occurrence in many human endeavors. The public would be wise to accept it as a part of these endeavors. For example, we accept that advertising is often biased and act accordingly. We tune out, we look for other sources (often from competitors), we view certain claims with skepticism. In the case of polling, a similar approach seems appropriate.

***Acknowledgment***

We thank Dustin Reishus of USC for his assistance in the computational aspects of this research.